\documentclass[aps,preprint,superscriptaddress]{revtex4}

\usepackage{graphicx}% Include figure files
\usepackage{dcolumn}% Align table columns on decimal point
\usepackage{bm}% bold math
\begin{document}

%Title of paper
\title{Climate Networks around the Globe are Significantly Effected by El Ni\~{n}o}

\author{K. Yamasaki}
  \affiliation{Tokyo University of Information Sciences, Chiba, Japan}
  \email{yamasaki@rsch.tuis.ac.jp}
\author{A. Gozolchiani}
\affiliation{
   Minerva Center and Department of Physics, Bar Ilan University, Ramat Gan, Israel.
}%

\author{S. Havlin}
\affiliation{
   Minerva Center and Department of Physics, Bar Ilan University, Ramat Gan, Israel.
}%

\date{\today}

\begin{abstract}
The temperatures in different zones in the world do
  not show significant changes due to El-Ni\~{n}o except when measured in a 
restricted area in the Pacific Ocean. We find, in contrast, that the
  dynamics of a climate network based on the same temperature records in 
  various geographical zones in the world is  significantly influenced
 by El-Ni\~{n}o. 
During El-Ni\~{n}o many links of the network are broken, 
and the number of surviving links comprises a specific and 
sensitive measure for El-Ni\~{n}o events. While during non El-Ni\~{n}o periods these links which represent correlations between temperatures in different sites are more stable, fast fluctuations of the correlations observed during El-Ni\~{n}o periods cause the links to break. 
\end{abstract}


\maketitle

Networks are often used to formulate the dynamics of complex systems
that are built from many interacting components (see e.g. ~\cite{acebron,hiroshi}). While in some cases the representation of the system as a
network is obvious and the nodes and links are identified directly
(e.g. cables connecting computers in a computer network)~\cite{barabasi}, there are cases in which the
process that couples the individual interacting components is more
complex and a link is guessed by tracking similarities in the dynamical behavior of two nodes~\cite{bootstrap}.

Even when the usual traces of dynamics of interacting nodes on a network, such as partial synchronization, clusters with correlated dynamics, oscillatory synchronization~\cite{tdelay_strogatz}, and phase slips~\cite{univar}, are evident, the mission of designing a generic tool that reliably extracts information about the network structure from measurements of the dynamics of nodes is still far from being accomplished.

In this Letter we develop a method for generating climate networks, which is suitable for tracking structural changes in these 
networks (dynamics \textbf{of} a network). These changes 
correspond, in our case, to strong climate changes due to El-Ni\~{n}o. 
We find that networks constructed from temperature measurements on 
different sites in the world are changed dramatically during El-Ni\~{n}o 
events in a similar way. These structural changes are seen 
even for geographical zones where the mean temperature is not affected by El-Ni\~{n}o.

\begin{figure}
\includegraphics[scale=0.45]{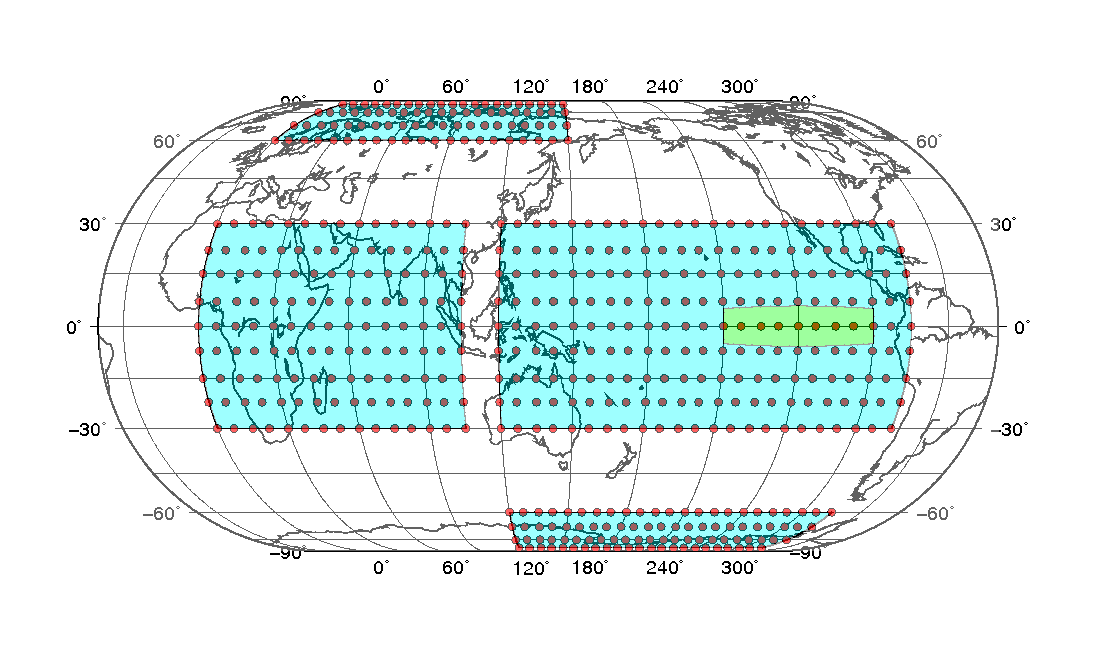}%
\caption{\label{fig_map}The four geographical zones used for building the four climate networks studied. The dots represent the nodes of the network. The rectangular geographical zone inside zone 1 shows the standard basin for which El-Ni\~{n}o effects on temperature and pressure is significantly observed (see Fig.~\ref{fig_survival}a)}
\end{figure}

We analyze daily temperature records taken from a grid 
(available at \cite{noaa}) in various geographical zones 
(shown in Fig.~\ref{fig_map}). 
To avoid the trivial effect of seasonal trends we 
subtract from each day's temperature, the yearly mean temperature 
of that day. Specifically, if we take the temperature signal of 
a given site in the grid to be $\widetilde{T}^{y}(d)$, 
where $y$ is the year and $d$ is the day (ranging from 1 to 365), 
the new filtered signal will be 
$T^{y}(d)=\widetilde{T}^{y}(d)-\frac{1}{N}\sum_{y}{\widetilde{T}^{y}(d)}$ 
(where $N$ is the number of years available in 
the record) \cite{fn1}.

We compute for a time shift $\tau\in\left[-\tau_{max},\tau_{max}\right]$ days for each pair of sites $l$ and $r$ on the grid, their cross correlation
function $X^{y}_{l,r}(\tau>0)\equiv\langle
T^{y}_{l}(d)T^{y}_{r}(d+\tau)\rangle_{d}$ and
$X^{y}_{l,r}(\tau\le 0)\equiv X^{y}_{r,l}(\tau>0)$. The correlation strength of the link is chosen to be $W^{y}_{l,r}=MAX(X^{y}_{l,r})/STD(X^{y}_{l,r})$, where
MAX and STD are the maximal value and the standard deviation of the absolute
value of $X^{y}_{l,r}$ in the range of $\tau$, respectively~\cite{fn2}. 
The time shift at which $X^{y}_{l,r}$ is maximal is defined as the time delay.
Up to here, the prescription is similar to other methods (see e.g.~\cite{tsonis} ).

From reasons that will become clear later, we are able to set a physical threshold $Q$ so that only pairs $l,r$ that satisfy $W^{y}_{l,r}>Q$, are regarded as significantly linked. Mathematically this can be represented by the Heaviside function 
$\Theta(x)$ as follows,

\begin{eqnarray}
\rho^{y}_{l,r}=\Theta(W^{y}_{l,r}-Q)
\label{eq:rho_def}.
\end{eqnarray}

Some of the elements of the
matrix $\rho$ may blink as a function of $y$, i.e., appear and disappear. 
Thus, even though there are
many pairs $l,r$ that their correlation values $W^{y}_{l,r}$ in a
specific year are above $Q$, some of these $W^{y}_{l,r}$ are sensitive
to the choice of the beginning of the period $y$, and to noise . In the
present work we choose to discard the  question of which pairs
comprise the static network, and concentrate in the structural changes of the network over time. Blinking links seem to be a signature of structural changes, so we distinguish between them and the more robust links, that are stable during larger time periods.

In the next step we examine if a currently existing link 
$\rho^{y}_{l,r}$ existed in the earlier periods of the network. To 
accomplish this we define a new matrix, which takes into 
account previous states of the links in the last $k$ states of the network. We define a new matrix $M^{y}_{l,r}$ which counts the number of times  a link appeared before continuously (without a blink):

\begin{eqnarray}
M^{y}_{l,r}=\sum_{n=0}^{y-1}\prod_{m=y-n}^{y}\rho^{m}_{l,r}
\label{eq:M_def}.
\end{eqnarray}

A link $\left(l,r\right)$ in the current network $\rho$ appeared $k$ times in a row before (including its current appearance) iff $M^{y}_{l,r}\ge k$. These links represent long lasting relations between temperature fluctuations in the zone. Counting them enables us to distinguish between the two qualitatively different groups of links, blinking links which are removed, and robust links which we include in the network. The number of links that exist in our network depends on $y$,

\begin{eqnarray}
n_{k}\left(y\right)=\sum_{l=0}^{N}\sum_{r=l+1}^{N}\Theta\left(M^{y}_{l,r}-k+1\right)
\label{eq:n_def}.
\end{eqnarray}

Where $k$ is the number of times a link has to survive in order to be
included in the network (according to Eq. (\ref{eq:M_def})), and $N$ is
the total number of links in the geographical zones. The summand in the
rhs of Eq. (\ref{eq:n_def}) represents the network matrix. In the current work we chose the $y$ resolution (the jumps between two subsequent dates represented by $y$) to be 50 days,  $k=5$ and
the threshold $Q=2$.

The results shown below are not sensitive to the choice of $k$. However, choosing too large $k$ values reduces the number of surviving links
significantly, and therefore eliminates much of the effect. Choosing too
small values of $k$, on the other hand, does not enable the elimination
of blinking links, and therefore causes $n_{k}\left(y\right)$ to be more
noisy, but the significant effect of breaking links is still evident.

We chose four representative zones around the globe, as shown in Fig.~\ref{fig_map}. 
Our networks are built from measurements of temperatures close to sea
level (networks $A_{j}$), and from measurements on a 500mb pressure
level (networks $B_{j}$), on a grid of $7.5^{o}$ resolution. The
measurements are taken for the years 1979-2006, for which 8 known
El-Ni\~{n}o events have occurred .

\begin{figure}
\begin{center}
\includegraphics[scale=0.08,angle=0]{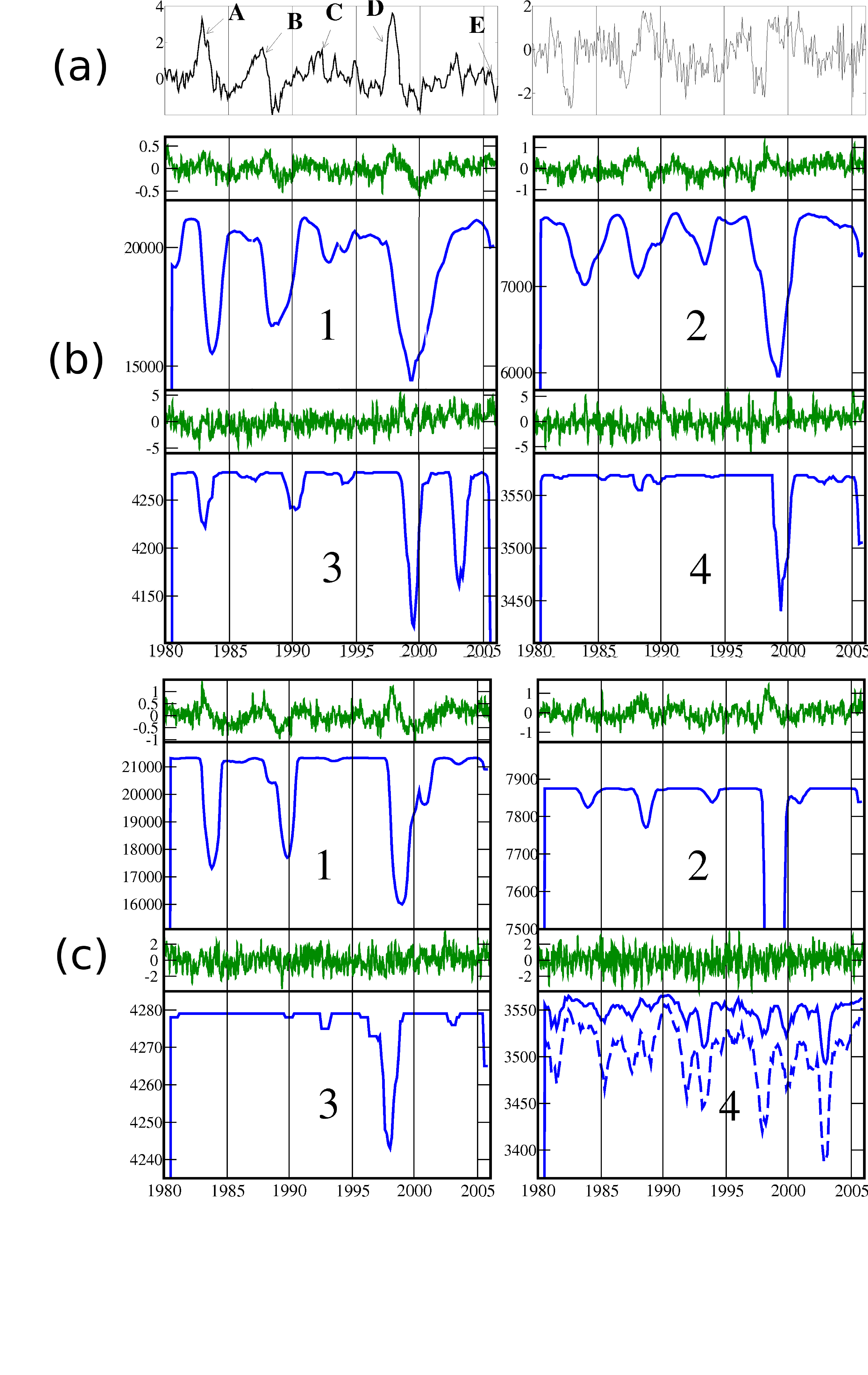} \\
\end{center}
\caption{(a) Mean sea surface temperature (left) in the standard basin shown in the rectangle inside $A_{1}$ in Fig.~\protect{\ref{fig_map}} (NINO3), and the difference in sea level pressure (right) between Tahiti and Darwin (SOI), both are standard indices for El-Ni\~{n}o (see e.g. \protect{\cite{dijkstra}}). (b) and (c) (color online) The upper curves represent the temperature anomaly series in zones (b)  $A_{1},A_{2},A_{3},A_{4}$ and (c)  $B_{1},B_{2},B_{3},B_{4}$. The lower curves present $n_{k}\left(y\right)$, namely, the number of links that survives in the network as a function of time, for these same zones. In zone $B_{4}$, the graph for $n_{k}\left(y\right)$ is completely flat, when $Q=2$. For this zone we show the cases of $Q=2.4$ (middle curve) and $Q=2.5$  (lower curve)~\protect{\cite{fn4}}}
\label{fig_survival}
\end{figure}

In Fig.~\ref{fig_survival}a we show the effect of El-Ni\~{n}o as 8 main
extreme values on the two standard El-Ni\~{n}o indices (based on temperature and pressure measurements), measured in the standard basin region inside zone 1 of Fig.~\ref{fig_map}  
\cite{fn5}.
In the top of each of the eight panels in  Fig.~\ref{fig_survival}b,c 
we show the mean
temperature anomaly over the whole corresponding zone defined in Fig.~\ref{fig_map}.
It is clearly seen that compared to Fig.~\ref{fig_survival}a the
El-Ni\~{n}o effect on mean temperature in all zones becomes very weak and almost cannot be detected  except at zones $A_{1}$ and $B_{1}$ (which
include the standard basin). In most zones, El-Ni\~{n}o effect on the temperatures are masked by noise of the same order. In mark contrast, when
measuring the number of links $n_{k}\left(y\right)$ in the climate
network we observe (in the bottom of Figs.~\ref{fig_survival}b,c) a significant effect of El-Ni\~{n}o in all four
zones. This is represented by the sharp fall of  $n_{k}\left(y\right)$
in most of the times of occurrence of El-Ni\~{n}o. In zones $A_{1},A_{2},A_{3}$ between 5 and 6 out of the 8 events can be observed while in zone $A_{4}$ (in the north) 
only the strongest El-Ni\~{n}o is seen clearly. 
Because of the close occurrence of the 3 events in the early 90's it
seems that two of the minima of $n_{k}\left(y\right)$ overlap. It is
also notable that in the zone surrounding the El-Ni\~{n}o basin (zone
$A_{1}$) the minimum due to the 2002 event cannot be inferred. However
in the $B_{1}$ (Fig.~\ref{fig_survival}c) network's $n_{k}\left(y\right)$ series the can be seen. In zone $A_{4}$ it is seen that even when the influence of El-Ni\~{n}o on the mean temperature in the zone is not seen at all,
the largest El-Ni\~{n}o event in 1997 as well as the event of 2005 can clearly be identified by a sharp
minimum in $n_{k}\left(y\right)$. 

In zone $B_{1}$ (Fig.~\ref{fig_survival}c) the main events are also clearly seen, but the 3 events in the early 90's
overlap completely. In network $B_{2}$ all events, except the 2002 event, can be observed. In network $B_{3}$ only the 1997 event is observed reliably. In network $B_4$ the traces of El-Ni\~{n}o can also be speculated if one sets the threshold to a higher value.  
\begin{figure}
\includegraphics[scale=0.3]{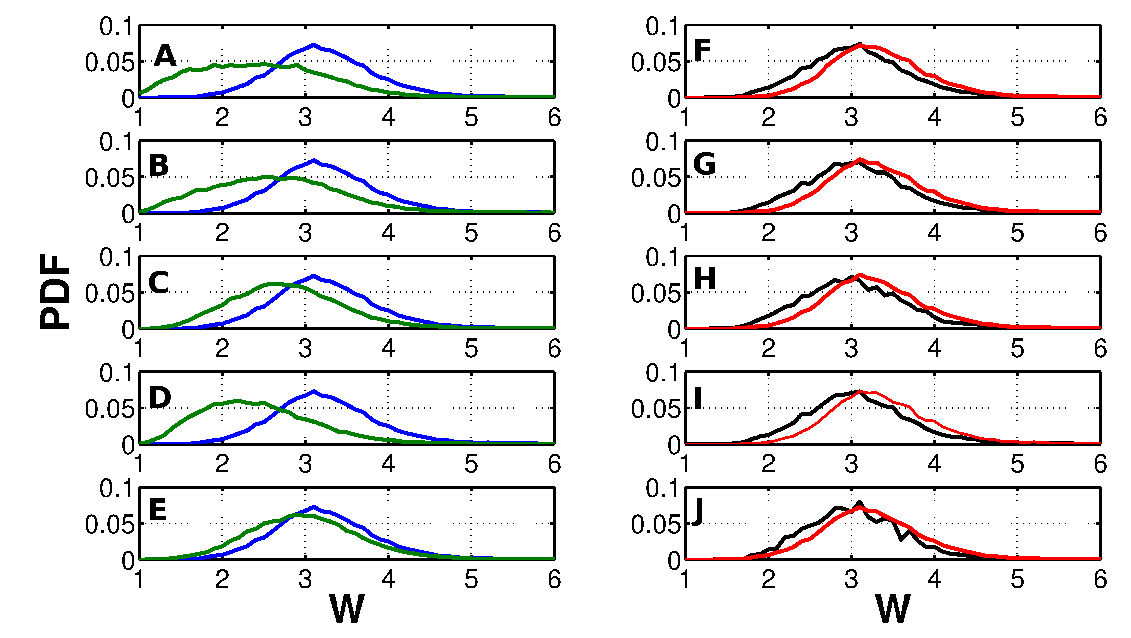}%
\caption{\label{fig_hist} (color online) A-E: The distribution of $W^{y}_{l,r}$ in zone 1 (see Fig.~\ref{fig_map}). The five rows correspond to the five El-Ni\~{n}o events pointed with an arrow in Fig.~\protect\ref{fig_survival}a, which are the most noticeable on zone 1. The curves with a peak to the right (blue) describe the distribution in a reference time, where the influence of El-Ni\~{n}o on the network is not seen. The curves with a peak to the left (green) describe the distribution for five El-Ni\~{n}o events. F-J: Two distributions of $W^{y}_{l,r}$ in the reference point (blue curves of graphs A-E). The curves slightly shifted to the left (black) are the distributions of the correlation values of the links that later on will disappear due to one of the El-Ni\~{n}o events , and the curves slightly shifted to the right (red) are the distribution of the ones that will remain connected in the considered event.}
\end{figure}

The choice of the threshold $Q=2$ is not arbitrary. When observing the probability density function of $W$ (Fig.~\ref{fig_hist}) it is clear that for non El-Ni\~{n}o time regimes, $W=2$ is actually the minimal value that exists. It therefore appears that choosing this
threshold  makes the network very sensitive to El-Ni\~{n}o events while
 remaining insensitive to other changes in climate. The reason is that
the distribution of $W^{y}_{l,r}$ tends to typical lower values of
$W^{y}_{l,r}$ during
El-Ni\~{n}o, as can be clearly seen in Fig.~\ref{fig_hist}(A-E). A
remarkable property of this softening is that the lower limit of the
distribution drops from being close to 2 to some significantly lower value. Changes in
climate around the world due to El-Ni\~{n}o events thus share a unified
property of the correlation pattern, which can be tracked in a reliable
way by the number of surviving links $n_{k}\left(y\right)$ in the climate network.

\begin{figure}
\includegraphics[scale=0.3]{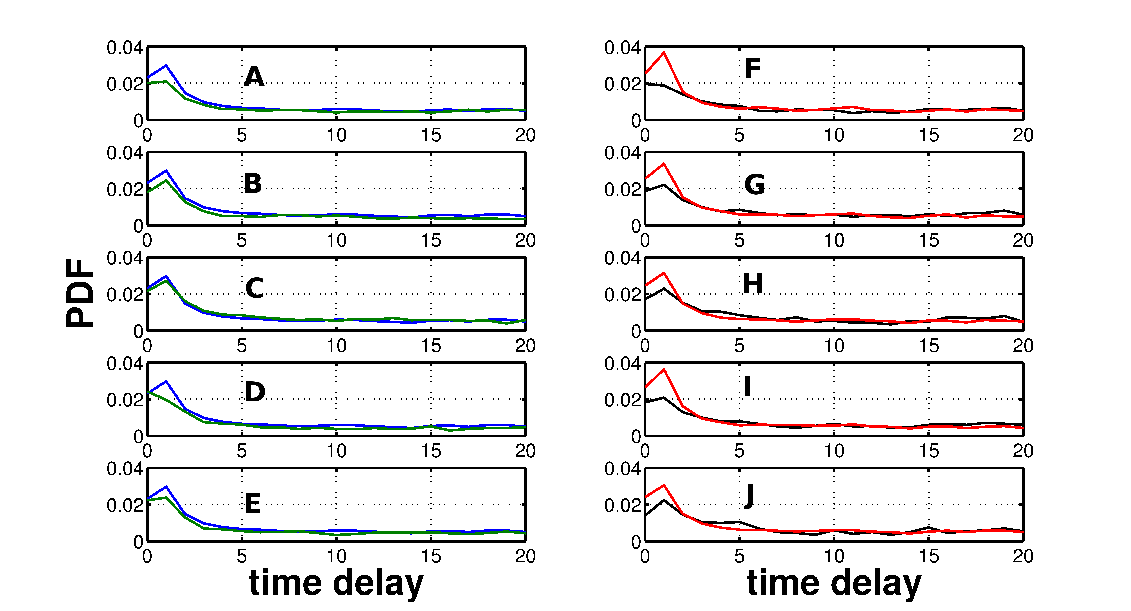}%
\caption{\label{fig_delays} (color online) Distribution of time delays. A-E, Lower (green) curves : The distribution of time delays of all links in the reference time, with no signs of El-Ni\~{n}o. Upper (blue) curves: Time delays distribution during five El-Ni\~{n}o episodes. F-J : Distribution of time delays of two groups of links during the reference time. Upper (red) curves - distribution of time delays of links that remain during El-Ni\~{n}o. Lower (black) curves- distribution of time delays of links that break during El-Ni\~{n}o.}
\end{figure}

How susceptible are the links that will break during El-Ni\~{n}o? As can
be seen in Fig. \ref{fig_hist}(F-J), the links in the fully connected
network that will later break have distribution of correlation strengths
very similar to that of the links that will survive. Only a slight tendency of
weaker links to break can be observed. However, when looking at the distribution of the time delays (Fig.~\ref{fig_delays}), there is a clear difference between links that survive and links that collapse. Links that collapse have a much broader distribution of time delays, and therefore have usually delays of more than a few days. Links with  shorter time delays have higher probability to survive.

An alternative approach that yields a similar effect can be obtained by counting the number of elements of $\rho^{y}_{l,r}$ that during the periods $\{y-2,y-1,y\}$ (two subsequent dates y,y+1 are separated by 50 days) posses for $\rho^{y}_{l,r}$  the sequence $\{1,0,1\}$ or the sequence $\{0,1,0\}$. Such a treatment which tests directly the relation between El-Ni\~{n}o and blinking of links yields results very similar to that given by $n_{k}\left(y\right)$. The blinking correlations events during El-Ni\~{n}o may arise from a gradual breaking of indirect paths between two nodes $l$ and $r$, or from a sensitivity to the choice of $Q$. On either case these events are a direct and unified mark of an aggressive structural change, that until now was described as various climate responses in various geographical zones. A more resolute theoretical problem of reconstructing an accurate network structure from experimental observations of its nodes' dynamics (in the current paper - the temperature ) remains largely open (see related discussions in~\cite{scc,atay,granger,perturb} ).

In summary, we have developed a method which enables one to follow large
changes over time of an underlying network structure by observations of fluctuations in the correlations between nodes. The method tracks blinking links that appear and
disappear in a short time, and assumes this behavior to be due to structural changes. Tracking the changes in the network of
temperatures in several zones in the world reveals a deep violent
response to El-Ni\~{n}o even in zones and heights where the mean
temperature level is not affected. The links that break during
El-Ni\~{n}o are mostly links that have large time delays.

\bibliography{correlation_networks}

\begin{thebibliography}{17}
\bibitem[{fn1()}]{fn1}
\bibinfo{note}{After filtering is finished, we can use overlapping year
  periods. From here on, \protect{$y$} represents the beginning date of a year
  period.}
\bibitem[{fn2()}]{fn2}
\bibinfo{note}{Division by the STD forces us to choose a wide span of $\tau$.
  Otherwise, significant peaks may have low values of $W^{y}_{l,r}$.
  Specifically, we chose $\tau_{max}=200$.}
\bibitem[{fn4()}]{fn4}
\bibinfo{note}{The date related to each point in the graph is the end of the
  $k$ (which equals 5 in our case) subsequent periods that we take into
  account. The time resolution (of $y$) for which we compute the network is 50
  days, which means that each point uses 250+365 days of information from its
  past. Because cross correlation shifts the data, we also use 200 days from
  the future of that point (which is our maximal shift).}
\bibitem[{fn5()}]{fn5}
\bibinfo{note}{This partial region is the standard region from which the
  El-Ni\~{n}o effect is determined and quantified.}
\end{thebibliography}

\end{document}